\documentclass[runinaddress,showpacs,twocolumn,prb]{revtex4}%
\usepackage{amssymb}
\usepackage{amsmath}
\usepackage{graphicx}
\usepackage{epstopdf}
\usepackage{amsfonts}
\usepackage{color}%
\providecommand{\U}[1]{\protect\rule{.1in}{.1in}}
\begin{document}
\title{Topological Anderson insulator phase in a Dirac-semimetal thin film}
\date{\today }

\author{Rui Chen}
\author{Dong-Hui Xu\footnote{donghuixu@hubu.edu.cn}}
\author{Bin Zhou\footnote{binzhou@hubu.edu.cn}}

\affiliation{Department of Physics, Hubei University, Wuhan 430062, China}

\begin{abstract}
{The recently discovered topological Dirac semimetal represents a new exotic
quantum state of matter. Topological Dirac semimetals can be viewed as three
dimensional analogues of graphene, in which the Dirac nodes are protected by
crystalline symmetry. It has been found that quantum confinement effect can gap out
Dirac nodes and convert Dirac semimetal to a band insulator. The band
insulator is either normal insulator or quantum spin Hall insulator
depending on the thin film thickness. We present the study of disorder
effects in thin film of Dirac semimetals. It is found that moderate Anderson
disorder strength can drive a topological phase transition from normal band insulator
to topological Anderson insulator in Dirac semimetal thin film. The numerical calculation based on
the model parameters of Dirac semimetal Na$_{3}$Bi shows that in the topological Anderson insulator phase
a quantized conductance plateau occurs in the bulk gap of band insulator,
and the distributions of local currents further confirm that
the quantized conductance plateau arises from the helical edge states induced by disorder.
Finally, an effective medium theory based on Born approximation fits the numerical data.}

\end{abstract}

\maketitle

\section{Introduction}

Two-dimensional (2D) massless Dirac fermions have been observed in graphene and on the surface of
three-dimensional (3D) topological insulators (TIs).\cite{Novoselov2004,Neto2009,Kotov2012,Moore2010,Hasan2010,Qi2011}
Recently, much attention has been attracted
to the Dirac semimetal which represents a new state of quantum matter, can
support 3D Dirac fermions with the linear dispersion in all three momentum
directions. The stable Dirac semimetals have been realized experimentally in two
crystalline materials, Na$_{3}$Bi and Cd$_{3}$As$_{2}$ compounds.\cite{Wang12PRB,Wang13PRB,Liu14Science,Liu14Natmat,Neupane2014} The
stability of the 3D Dirac points in these materials requires additional
crystalline symmetries other than the time-reversal symmetry and inversion
symmetry.\cite{Yang2014} Dirac semimetals show many interesting physics properties, such as
spin polarized double Fermi arc surface states and chiral
anomaly in the presence of parallel electric and magnetic fields.\cite{Xu2015,Xiong2015}

Finite size effect plays an important role in the system with size comparable
to the quasiparticle wavelength. In 2DTIs, the states on the opposite edges
can couple with each other to open a gap in the gapless edge energy spectrum
due to finite size effect.\cite{Zhou08PRL} While for the thin film of 3DTIs,
depending on the thickness, it may become a quantum spin Hall (QSH) insulator or a
trivial band insulator.\cite{Linder09PRB,Zhang10NP,Lu10PRB,Liu10PRB,Imura12PRB} Similar properties
are also proposed in Dirac semimetal materials Na$_{3}$Bi and Cd$_{3}$As$_{2}$, in which the quantum confinement effect can induce a
periodically modulated gap at Dirac node.\cite{Wang12PRB,Wang13PRB,Xiao2015,Pan2015}
Whether a finite size gap is inverted or not is determined by the thickness of
Dirac semimetal thin film. Thus, Dirac semimetals Na$_{3}$Bi and Cd$_{3}$As$_{2}$
should cross over between trivial and nontrivial 2D insulators oscillatorily as a function
of Dirac semimetal thin film thickness.\cite{Wang12PRB,Wang13PRB}

It is well known that disorder has a major impact upon the transport
properties of low-dimensional electronic systems.\cite{Lee58RMP} In 2009, Li
\emph{et al}. reported that the on-site Anderson type disorder can
trigger a phase transition from topologically trivial phase to QSH phase
with quantized edge conductance.\cite{LiJ09PRL} This disorder-induced topological
insulator was named as topological Anderson insulator (TAI).\cite{LiJ09PRL} This TAI was
quickly confirmed by the numerical simulations and Born
approximation analysis in which the disorder-induced
topological phase transition can be understood by the renormalization of Dirac
mass and chemical potential.\cite{Jiang09PRB,Groth09PRL} TAI has been investigated in more related models,
such as Haldane model, Kane-Mele model and 3DTI model.\cite{Xing11PRB,Orth16Scirep,GUO10PRL,Guo11PRB,Guo10PRB,Jiang16cpb}
It is indicated that TAI
is dependent on the type of disorder, for example, TAI phase is
absent for the magnetic disorder and spatially correlated
disorder.\cite{Song12PRB,Girschik13PRB} On the other hand, gapped and gapless topological phases and transport characteristics in layered structures and thin films have been studied recently in the presence and absence of disorder.\cite{Burkov11PRL,Kobayashi15PRB,Yoshimura16PRB} It is interesting to note that Kobayashi \emph{et al}. revealed the TAI phase in topological insulator nanofilms by numerical calculations.\cite{Kobayashi15PRB}

In this paper, we investigate the interplay of disorder and quantum confinement
 in Dirac semimetal thin film. It is found
that the Anderson disorder can induce a topological phase transition in this
system. Topological Anderson insulating phase emerges by tuning the strength
of disorder. By combining numerical simulation based on the recursive Green's
function and Born approximation with
the model parameters of Dirac semimetal Na$_{3}$Bi, we show that when the on-site disorder is
introduced to the insulating phase of Dirac semimetal with spin Chern number
$C_{s}=0$, the TAI phase with a quantized conductance
plateau ($\sigma=2e^{2}/h$) appears in a certain range of disorder strength.
Inspection of the nonequilibrium local current distribution further confirms
that this quantized conductance plateau arises from the helical edge states induced by disorder.

\section{Model}

We start with a generic low-energy effective model derived from first
principles results of Dirac semimetals A$_{3}$Bi(A=Na,K,Rb)
and Cd$_{3}$As$_{2}$.\cite{Wang12PRB,Wang13PRB} In the basis of $\left\vert S_{\frac
{1}{2}},\frac{1}{2}\right\rangle ,\left\vert P_{\frac{3}{2}},\frac{3}%
{2}\right\rangle ,\left\vert S_{\frac{1}{2}},-\frac{1}{2}\right\rangle
,\left\vert P_{\frac{3}{2}},-\frac{3}{2}\right\rangle $, the Hamiltonian of
effective model can be expressed as
\begin{equation}
H\left(  \mathbf{k}\right)  =\epsilon_{0}\left(  \mathbf{k}\right)  +%
\begin{pmatrix}
M\left(  \mathbf{k}\right)  & Ak_{+} & 0 & 0\\
Ak_{-} & -M\left(  \mathbf{k}\right)  & 0 & 0\\
0 & 0 & M\left(  \mathbf{k}\right)  & -Ak_{-}\\
0 & 0 & -Ak_{+} & -M\left(  \mathbf{k}\right)
\end{pmatrix}
, \label{eq1}%
\end{equation}
where $\left\vert S_{\frac{1}{2}},\pm\frac{1}{2}\right\rangle $ are the
conduction $s$ state and $\left\vert P_{\frac{3}{2}},\pm\frac{3}%
{2}\right\rangle $ are the heavy-hole $p$ state, $\epsilon_{0}\left(
\mathbf{k}\right)  =C_{0}+C_{1}k_{z}^{2}+C_{2}k_{\Vert}^{2}$, $M\left(
\mathbf{k}\right)  =-M_{0}+M_{1}k_{z}^{2}+M_{2}k_{\Vert}^{2}$, $\mathbf{k}%
_{\Vert}=\left(  k_{x},k_{y}\right)  $, and $k_{\pm}=k_{x}\pm ik_{y}$. There
are two Dirac nodes located at $(0,0,\pm k_{D})$ with $k_{D}=\sqrt{M_{0}%
/M_{1}}$.

In the following calculations, we will take the model parameters of Dirac semimetal Na$_{3}%
$Bi obtained from the first-principles calculations,\cite{Wang12PRB} namely $C_{0}=-63.82$
meV, $C_{1}=87.538$ meV$\cdot$nm$^{2}$, $C_{2}=-84.008$ meV$\cdot$nm$^{2}$,
$M_{0}=86.86$ meV, $M_{1}=106.424$ meV$\cdot$nm$^{2}$, $M_{2}=103.610$
meV$\cdot$nm$^{2}$, and $A=245.98$ meV$\cdot$nm. We discretize the effective Hamiltonian on a
3D simple cubic lattice and set the lattice constants as
$a_{x}=a_{y}=a_{z}=a=0.5$ nm.

\begin{figure}[hptb]
\includegraphics[width=8cm]{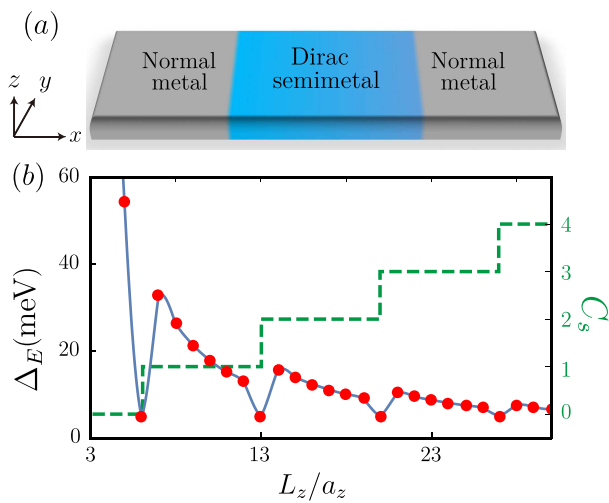} \caption{(Color online) (a) Schematic
illustration of metal-Dirac semimetal-metal setup. (b) Confinement-induced
bulk gap $\Delta_E$ (blue solid line with red dots) and the spin Chern number $C_s$ (green
dashed line) versus thin film thickness $L_{z}$. }%
\label{fig1}%
\end{figure}

Now let us consider a thin film of Dirac semimetal confined along $z$
direction with thickness $L_{z}$. Due to quantum confinement effect, the
energy band along $z$ direction will be quantized into separated levels. The
confinement-induced band gap as a function of thickness is presented in Fig.
1(b). Using quantum well approximation, the following term can be obtained for
each subband $n$ \cite{Xiao2015,Pan2015}
\begin{equation}
M\left(  \mathbf{k}\right)  \rightarrow M\left(  n,\mathbf{k}_{\Vert}\right)
=\mathcal{M}_{n}+M_{2}k_{\Vert}^{2},
\end{equation}
where $\mathcal{M}_{n}=-M_{0}+M_{1}\left(  n\pi/L_{z}\right)  ^{2}$ is the
mass parameter of these subbands. For each subband $n$, the Hamiltonian (\ref{eq1}) is
similar to the BHZ model Hamiltonian describing the QSH
system in HgTe/CdTe quantum wells.\cite{Bernevig06Science}

\begin{figure*}[phtb]
\includegraphics[width=17cm]{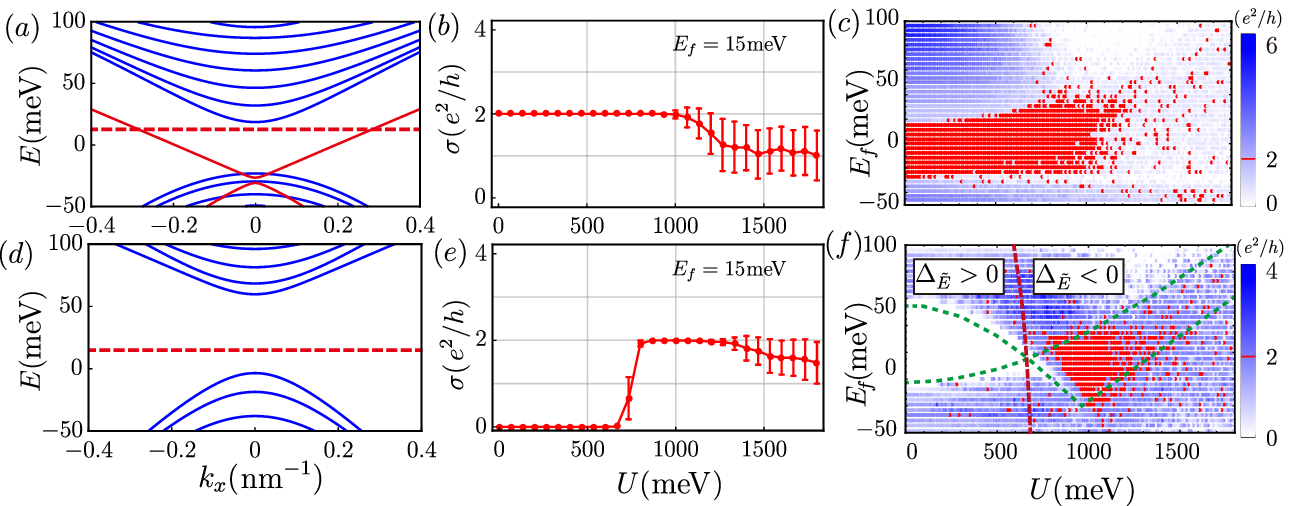} \caption{(Color online) (a) Energy
spectrum of Dirac semimetal thin film with open boundary conditions along $y$ and $z$ directions and periodic boundary condition along $x$ direction. Here, the sizes in the $y$ and $z$ directions are taken as $L_{y}=80a_{y}$ and $L_{z}=7a_{z}$, respectively. The blue curves are the bulk subbands and the red lines correspond to the surface
states. (b) The disorder averaged conductance as a function of disorder
strength $U$. The error bars show standard deviation of the conductance for
500 samples. The Fermi level is $E_{f}=15$ meV. (c) Phase diagram
showing the conductance as a function of disorder strength $U$ and Fermi
energy $E_{f}$. Each data point corresponds to a single realization of the
disorder potential. The data points marked in red represent the value of conductance within the interval $(2-\epsilon)e^2/h<\sigma <(2+\epsilon)e^2/h$, and we take $\epsilon=0.005$. The other panels (d), (e) and (f) are the same with (a), (b)
and (c), except that $L_{z}=5a_{z}$. The red and green dashed lines in (f)
correspond to the phase boundary defined as $\Delta_{\tilde{E}}=0$ and
$\tilde{E}_{2}<\tilde{E}_{f}<\tilde{E}_{1}$. In our numerical simulations, the system size is chosen as $L_z=7a_z$, $L_{y}=80a_{y}$ and
$L_{x}=300a_{x}$ in (b) and (e), and $L_z=5a_z$, $L_{y}=80a_{y}$ and
$L_{x}=150a_{x}$ in (c) and (f).}%
\label{fig2}%
\end{figure*}

To characterize the system, we will calculate the spin Chern number $C_{s}$
within quantum well approximation.\cite{Sheng06PRL,Prodan09PRB,Li10PRB} It
has been shown that the spin Chern number and $Z_{2}$ topological invariant
would yield the same classification by investigating topological properties of
time-reversal invariant systems.\cite{Fukui07PRB} The spin Chern
number is defined as\cite{Yang14EPL}
\begin{equation}
C_{s}  =\frac{1}{2}\sum_{n}{(C_{\uparrow}^{n}-C_{\downarrow}^{n})},
\end{equation}
where $C_{\uparrow}^{n}$ and $C_{\downarrow}^{n}$ are the valence band Chern
number of the spin up and down blocks of the $n$-th subband, respectively. The results are
shown in Fig. \ref{fig1}(b), in which the green
dashed line depicts the spin Chern number.
When the thickness of this system is small ($L_{z}<6a_{z}$), the spin Chern
number keeps zero, indicating that the system is actually a normal insulator.
While if the thickness exceeds a critical value ($L_{z}\approx6a_{z}$), the
energy gap closes and reopens, and a nonzero spin Chern number $C_{s}=1$ is
obtained, which means that a topological phase transition from the normal insulator
to the QSH state has happened. With further increasing $L_{z}$,
this system shows more than one Kramer's pair of helical edge states,
characterized by higher spin Chern numbers with $C_{s}=2,3,\cdots$. The spin Chern number
is exactly equal to the number of the pair of one-way helical edge states. The odd spin Chern number corresponds to
the QSH phase, while the even spin Chern number corresponds to the normal insulator. Thus this system crosses
over between trivial and nontrivial 2D insulators oscillatorily as a function
of thin film thickness. The result given by the spin Chern number is consistent with
the previous results.\cite{Wang12PRB,Wang13PRB}

To better understand the different phase in Dirac semimetal thin film tuned by
finite size effect, we plot the energy band spectrum for thickness
$L_{z}=7a_{z}$ and $L_{z}=5a_{z}$ in Fig. \ref{fig2}(a) and (d), which
correspond to the QSH state and normal insulating state,
respectively. For the case of $L_{z}=5a_{z}$, since all the subbands are
trivial, there is no edge state on the side surfaces of Dirac semimetal. While
for $L_{z}=7a_{z}$, one subband becomes inverted, as a result, there is a pair
of helical edge states on the side surfaces, as indicated by the red
curves. For a thicker sample with $L_{z}=15a_{z}$, there will exit two pairs of
helical edge states, corresponding to the spin Chern number $C_{s}=2$.

\section{Numerical simulation}
Now, we will investigate the transport properties
of the Dirac semimetal thin film in the presence of disorder by using the
Landauer-B\"uttiker-Fisher-Lee formula
\cite{Landauer70PhiMag,Buttiker88PRB,Fisher81PRB} and the recursive Green's
function method.\cite{MacKinnon85PhysBCM,Metalidis05PRB} The linear conductance can be obtained by $\sigma=(e^2/h)T$, where $T=$ Tr$\left[ \Gamma_L G^r \Gamma_R G^a \right]$ is the transmission coefficient. The linewidth function $\Gamma_{\alpha}(E_f)=i\left[ \Sigma_{\alpha}^{r}-\Sigma_{\alpha}^{a}\right]$ with $\alpha=L,R$, and the Green's functions $G^{r/a}(E_f)$ are calculated from $G^{r}(E_f)=\left[ G^a(E_f)\right]^\dag=\left[EI-H_C-\Sigma_{L}^{r}-\Sigma_{R}^{r} \right]^{-1}$, where $E_{f}$ is the Fermi level, $H_C$ is the Hamiltonian matrix of the central scattering region, and $\Sigma_{L,R}^{(r/a)}$ are the retarded (advanced) self-energy due to the device leads.

We adopt a rectangular thin film sample of size $L_{x}\times L_{y}\times L_{z}$ and two
semi-infinite metal leads connected to the sample along the $x$ direction as
shown in Fig. \ref{fig1}(a). To avoid redundant scattering from mismatched interface between the leads and central scattering region, the two leads are also modeled by clean Dirac semimetal thin films. In our numerical simulations, we set the chemical potentials of leads $\mu_{L,R}=100$ meV to guarantee a high density of states. We introduce the Anderson type disorders to the
central scattering region through random on-site energy with a uniform distribution
within $[-U/2,U/2]$, with the disorder strength $U$. This setup allows us to calculate the longitudinal conductance $\sigma$ and the distribution of local current.

Fig. \ref{fig2}(b) and
\ref{fig2}(e) show the conductance $\sigma$ of Dirac semimetal thin film with
different thickness ($L_{z}=7a_{z}$ and $5a_{z}$) versus disorder strength. The
Fermi level $E_{f}=15$ meV is chosen inside the band gap caused by
confinement effect. We found that the conductance keeps quantized without any
fluctuations if the disorder strength $U$ is smaller than $U_{c}=1$ eV when the
system is in the QSH regime ($L_{z}=7a_{z}$). The conductance plateau
collapses when the disorder is strong enough. The quantized conductance
plateau manifests the robustness of helical edge states to the weak disorder.
Such observation agrees with the previous result in the literatures.
\cite{Kane05PRL1,Kane05PRL2,Sheng05PRL,Taskin12PRL} While intriguing
phenomena emerge in the normal state ($L_{z}=5a_{z}$). Initially, the
conductance is zero since there is no electronic state inside the band gap for
a normal insulator. With the increasing of the disorder strength, the
conductance becomes finite and reaches to a quantized value ($\sigma=2e^{2}%
/h$), then the conductance maintains at this value for a certain range before
eventually decreases. The absence of fluctuation in the quantized conductance plateau denotes that
 the quantized conductance is contributed by the helical edge states, indicating the topological nature of the TAI phenomena.

Fig. \ref{fig2}(c) and \ref{fig2}(f) show the phase diagram obtained by
numerical simulations. Each point corresponds to a single realization of the disorder potential, which turns out to be sufficient for determining the region of TAI phase. We find that this quantized region appears not only in
the band gap, but also occurs in a small region of the conduction bands
$(L_{z}=7a_{z})$ and the valence bands $(L_{z}=5a_{z})$.

\begin{figure}[htpb]
\includegraphics[width=8cm]{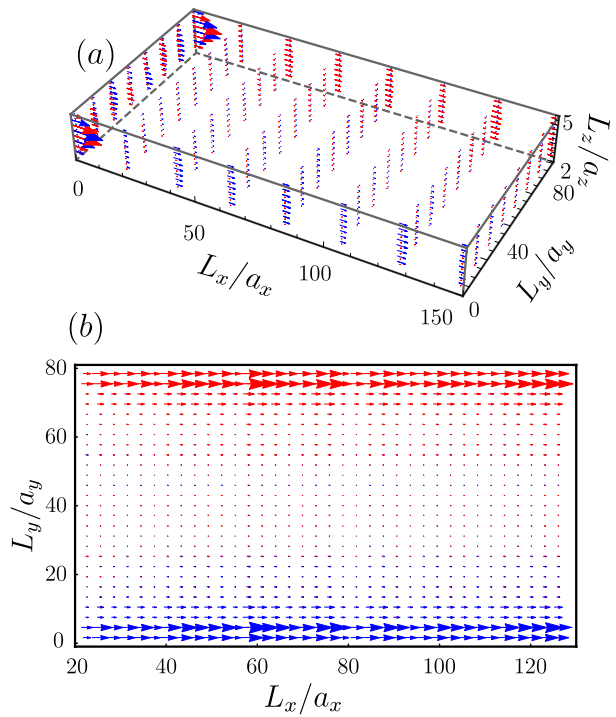} \caption{(Color online) (a) The averaged
nonequibrium local current distribution for the TAI phases of the Dirac semimetal
thin film with thinkess scale $L_{z}=5a_{z}$, Fermi level $E_{f}=15$ meV and the
disorder strength $U=1000$ meV in a two terminal setup. The red and blue arrows
correspond to the current of the spin-up and spin-down component, respectively. The arrow size
means the current strength. (b) The cross section of (a) with $L_{z}=3a_{z}$.
}%
\label{fig3}%
\end{figure}

To substantiate the
assertion that the quantized conductance plateau originates from the robust
edge states, we study the nonequilibrium local current distribution between neighboring sites $\bf{i}$ and $\bf{j}$ from the formula\cite{Jiang09PRB}
\begin{equation}
J_{\bf{i}\rightarrow\bf{j}}=\frac{2e^2}{h}\text{Im}
\left[\sum_{\alpha,\beta}{H_{\bf{i} \alpha,\bf{j}\beta}G^n_{\bf{j}\beta,\bf{i}\alpha}(E_f)}\right]
\left(V_L-V_R \right)\text{,}
\end{equation}
where $G^n(E_f)=G^r\Gamma_L G^a$ is electron correlation function. To calculate the local current distribution, a small external bias $V=V_L-V_R$ is applied longitudinally between the two terminals,
where $V_L$ and $V_R$ describe the voltages of the left and right leads. We assume the electrostatic potential in the left lead to be 1 meV and zero in the right lead. The electrostatic potential in the central part is $\phi(i_x)=(L_x-i_x+1)/(L_x+1)$ meV, where $i_x$ is the site index along the $x$-direction and $1\leq i_x \leq L_x$. Then, the electric field is uniformly distributed in the central device region. As shown in Fig. \ref{fig3}, the red and blue arrows correspond to the strength of local currents of the upper and lower blocks, respectively.
Apparently, the spin-up and spin-down local currents are localized at the two different side surfaces, respectively.
The QSH insulator is characterized by the helical edge states shown in Fig. 3.

\section{Born approximation}

To corroborate the physical interpretation of numerical simulation, we analyze
the present model within an effective medium theory based on the Born
approximation in which high order scattering processes are neglected.
\cite{Groth09PRL} In the self-consistent Born approximation, the self-energy
$\Sigma$ for a finite disorder strength is given by the following integral
equation%
\begin{equation}
\Sigma=\frac{U^{2}}{12}\left(  \frac{a}{2\pi}\right)  ^{2}\int_{1BZ}%
d\mathbf{k}_{\Vert}\frac{1}{E_{f}-H\left(  \mathbf{k}_{\Vert}\right)  -\Sigma
}\text{,}%
\label{Born}
\end{equation}
where $H\left(  \mathbf{k}_{\Vert}\right)  $ is the model Hamiltonian of
Dirac semimetal with the confinement imposed along $z$ direction. The
coeffcient 1/12 originates from the variance $\left\langle U^{2}\right\rangle
=U^{2}/12$ of a random variable uniformly distributed in the range $\left[
-U/2,U/2\right] $. This integration is
over the first Brillouin zone (1BZ). We will use the lowest-order Born
approximation, which means setting $\Sigma=0$ on the right hand side of Eq. (\ref{Born}).

When the confinement is
imposed along $z$ direction with $L_{z}=5a_{z}$, by calculating the bottom of
the conduction band $\tilde{E}_{1}$ and the top of the valence band $\tilde
{E}_{2}$ of the renormalized Hamiltonian $(\tilde{H}=H+\Sigma)$ as a function
of $E_{f}$ and $U$, we obtain the red and green dashed curves in Fig.
\ref{fig2}(f). The red curve denotes the phase boundary line $\tilde{E}%
_{1}-\tilde{E}_{2}=\Delta_{\tilde{E}}=0$, which means the gap closing. The
region inside the green dashed lines is determined by $\tilde{E}_{2}<\tilde
{E}_{f}<\tilde{E}_{1}$, which is another necessary condition for the TAI
phenomena, i.e., the renormalized Fermi level must be inside the renormalized
energy gaps. We find that the results based on Born approximation fit well with the previous numerical calculations, which means that disorder has a renormalized
effect on bulk band gap $\Delta_E$ and the Fermi level $E_{f}$, leading to
the TAI phenomena in Dirac semimetal thin films.

To demonstrate the disorder-induced topological phase transition, the spin
Chern number $\tilde{C}_{s}$ of the renormalized Hamiltonian $\tilde{H}$ is
also evaluated for various $U$ and $E_{f}$ by using the quantum well
approximation. It is found that the topological mass $\mathcal{M}_{n}$ of each
quantum well modes are renormalized by disorder. An approximate analytic
solution of the effective mass $\tilde{\mathcal{M}}_{n}$ is given by\cite{Groth09PRL}%
\begin{equation}
\tilde{\mathcal{M}}_{n}=\mathcal{M}_{n}-\frac{U^{2}a^{2}}{48\pi}%
\frac{M_{2}}{M_{2}^{2}-C_{2}^{2}}\ln\left\vert \frac{M_{2}^{2}-C_{2}^{2}%
}{E_{f}^2-\mathcal{M}_{n}^{2}}\frac{\pi^{4}}{a^{4}}\right\vert
\text{,}%
\end{equation}
showing that the disorder correction to the topological mass $\mathcal{M}_{n}$
is negative, provided by $M_{2}>0$ and $\left\vert M_{2}\right\vert
>\left\vert C_{2}\right\vert $. For a clean system with $L_{z}=5a_{z}$, all of
the subsystems are trivial, and this system is a normal insulator with
$C_{s}=0$. For a certain disorder strength (about 750 meV), the first quantum
well mode is inverted with a negative mass $\tilde{\mathcal{M}}_{1}<0$, and this
system turns to QSH insulator with $\tilde{C}_{s}=1$.

\section{Conclusion}

In this paper, we investigate disorder-induced TAI in Dirac semimetal thin
films. We observe a transition from a trivial insulating phase to a QSH
state at a finite disorder strength. We present the phase diagram as a
function of the disorder strength and the Fermi level, which is in accordance
with the result obtained by the effective medium theory based on the Born
approximation. We also plot the nonequilibrium local current
distribution, which further confirms that the quantized
conductance plateau in the TAI phase arises from the helical edge states induced by disorder.

\section*{Acknowledgments}

This work was supported by the National Natural Science Foundation of China
(Grant No. 11274102), the Program for New Century Excellent Talents in
University of Ministry of Education of China (Grant No. NCET-11-0960), and
Specialized Research Fund for the Doctoral Program of Higher Education of
China (Grant No. 20134208110001).

\end{document}